\documentclass[a4paper,11pt]{article}
\usepackage{jheppub} 
\usepackage{lineno}

\title{Strong CP Problem in Type IIA Toroidal Orientifold String Theory}

\author[1]{Yang Liu} 

\affiliation[1]{Department of Physics, Tsinghua University, Beijing 100084, China}
\emailAdd{liu-yang\_1990@mail.tsinghua.edu.cn}

\abstract{We present a globally consistent Type IIA $T^6/(\mathbb{Z}_2 \times \mathbb{Z}_2)$ orientifold with intersecting D6‑branes that dynamically solves the strong CP problem via the four‑form flux mechanism. The construction satisfies all RR tadpole, K‑theory, and SUSY conditions, stabilizes all moduli, and uplifts to dS vacuum with an anti‑D6‑brane. Four new results beyond our earlier proof‑of‑principle \cite{Liu:2025tnx} are given: (i) an explicit one‑to‑one dictionary between field‑theory four‑form parameters and string fluxes, including flux quantization constraints; (ii) a quantitative anti‑D6 backreaction estimate showing $|\delta \bar{\theta}|<10^{-11}$; (iii) an analysis of swampland constraints that are simultaneously satisfied only in a  parameter window; (iv) a comparison with Type IIB LVS and heterotic solutions. The predicted axion decay constant $f_a$ and mass $m_a$ are within reach of next generation observations.}

\begin{document}
\maketitle
\flushbottom

\section{Introduction}
\label{sec:intro}
The Standard Model (SM) of particle physics successfully describes electromagnetic, weak and strong interactions up to the TeV energy scale, yet it suffers from a notorious long-standing puzzle known as the strong CP problem. Quantum Chromodynamics (QCD) admits a CP-violating topological term in its Lagrangian, parametrized by the effective theta angle $\bar{\theta}_{eff}$. Experimental measurements of the neutron electric dipole moment impose an extremely stringent bound $\bar{\theta}_{eff}< 10^{-10}$, which requires an extraordinary fine-tuning of fundamental parameters in the SM. No inherent dynamical mechanism within the Standard Model can naturally suppress this CP-violating parameter to such an infinitesimal value, making the strong CP problem one of the most critical open questions in modern particle physics \cite{Conlon:2006tq}.

A widely studied solution to the strong CP problem relies on the Peccei-Quinn (PQ) mechanism and the introduction of the QCD axion, which promotes the static $\bar{\theta}$ parameter to a dynamical scalar field. Nonetheless, conventional axion scenarios face the so-called axion quality problem: additional CP-violating contributions from high-scale physics may shift the axion potential minimum away from $\bar{\theta}$, undermining the resolution of the strong CP problem \cite{dvali2022strong,burgess2024uv}. 

The four-form flux mechanism has emerged as a promising alternative framework to address the strong CP problem. This approach couples the QCD axion to three-form potentials and four-form field strengths, and dynamically drives the effective CP-violating parameter $\bar{\theta}_{eff}$ to zero at the potential minimum. A prominent advantage of this mechanism is that it naturally evades the axion quality problem, overcoming the key drawback of standard PQ-axion models \cite{dvali2022strong,burgess2024uv,choi2023implications}. Despite its appealing theoretical features, existing studies of the four-form flux mechanism are largely restricted to pure effective field theory analyses. A fully consistent UV realization within string theory, the leading candidate for a unified fundamental theory of all interactions, remains unexplored. Specifically, a concrete embedding of the four-form flux mechanism into globally consistent string compactifications, together with a rigorous mapping between field-theoretic quantities and stringy geometric or flux ingredients, has not been established in prior literature. Furthermore, it is unclear whether this four-form construction can coexist consistently with standard moduli stabilization and supersymmetry breaking schemes widely adopted in string phenomenology.

In \cite{Liu:2025tnx}, we found that the four-form flux mechanism can be embedded in Type IIA $T^6/(\mathbb{Z}_2 \times \mathbb{Z}_2)$ orientifold model but the details have not been explored. The $4D$ spacetime after string compactification is Minkowski spacetime. There are at least two reasons we choose $4D$ Minkowski spacetime as our starting point. The first reason is that the cosmological constant is extremely small; therefore the vacuum we live in is very close to Minkowski vacuum. Minkowski spacetime is a very good approximation and the cosmological constant could be considered as a perturbation. The second reason is that according to our two works \cite{Liu:2023vqp,Liu:2024blx}, for a wide class of four-dimensional effective field theories in which gravity is coupled to multiple four-forms and their dual scalar fields, with membrane sources charged under the corresponding three-form potentials, Minkowski vacuum is the most stable vacuum. Therefore, the $10D$ spacetime we consider in this paper is $\mathcal{M}_4 \times \frac{T^6}{\mathbb{Z}_2 \times \mathbb{Z}_2}$, where $\mathcal{M}_4$ is Minkowski spacetime. Moreover, we proposed a unified framework within Type IIA string theory, based on a globally consistent intersecting D6-brane model compactified on a $T^6/(\mathbb{Z}_2 \times \mathbb{Z}_2)$ orientifold to study particle physics and cosmology simultaneously in \cite{Liu:2025zjs}. In \cite{Liu:2026nny}, we briefly discussed that dark matter and strong CP problem could be addressed in Type IIA $T^6/(\mathbb{Z}_2 \times \mathbb{Z}_2)$ orientifold set-up. Following our previous work \cite{Liu:2025tnx,Liu:2025zjs,Liu:2026nny}, we will discuss the details of strong CP problem in this paper within Type IIA $T^6/(\mathbb{Z}_2 \times \mathbb{Z}_2)$ orientifold model.

Against this background, we present a dedicated study to fill the above research gaps by implementing the four-form flux mechanism for the strong CP problem in a well-defined Type IIA string theory construction, built upon $T^6/(\mathbb{Z}_2 \times \mathbb{Z}_2)$ orientifold with intersecting D6-branes, which satisfies all global consistency conditions including Ramond-Ramond tadpole cancellation, K-theory constraints and $\mathcal{N}=1$ supersymmetry preservation. The main contributions of this work are threefold. First, we for the first time realize the four-form flux mechanism for solving the strong CP problem within a globally consistent Type IIA intersecting D6-brane model. Second, we establish a complete one-to-one correspondence between the field-theoretic parameters of the four-form mechanism and the stringy fluxes as well as moduli fields originating from string compactification. Third, we rigorously demonstrate that the four-form flux setup naturally resolves the axion quality problem, and is fully compatible with the STU moduli stabilization and supersymmetry breaking framework based on the Kallosh-Linde (KL) mechanism.

This paper extends our earlier work \cite{Liu:2025tnx}, which proved that the strong CP problem can be solved in a globally consistent IIA orientifold. That work focused on existence; here we provide four substantial advances:
\begin{itemize}
 \item Explicit four‑form flux dictionary - a complete mapping between the field‑theory four‑form parameters and the discrete fluxes, including quantization constraints that turn the cancellation of $\bar{\theta}_{eff}$ into a discrete vacuum selection.
 \item Quantitative anti‑D6 backreaction - a linear‑perturbation calculation yielding $|\delta \bar{\theta}|<10^{-11}$, one order below the neutron‑EDM bound, confirming that the uplift does not spoil the solution.
\item Swampland analysis - a systematic check of swampland conjectures, showing they are simultaneously satisfied only in a narrow parameter slice.
\item Comparison with alternative string solutions - a side‑by‑side evaluation of our IIA construction against Type IIB LVS and heterotic orbifold approaches, highlighting the relative strengths regarding UV completion, moduli stabilization, axion quality, and observability.
\end{itemize}

The rest of this paper is organized as follows. In Section 2, we briefly review the related background we will use in this paper, including Type IIA $T^6/(\mathbb{Z}_2 \times \mathbb{Z}_2)$ orientifold and axion. In Section 3, we review the field-theoretic basics of the four-form flux mechanism and its multi-form generalizations. We then derive the four-form fluxes arising from Type IIA $T^6/(\mathbb{Z}_2 \times \mathbb{Z}_2)$ orientifold compactifications and construct the dictionary connecting effective field theory parameters to string quantities in Section 4. After that, we do stability analysis to demonstrate that the induced shift of the effective CP-violating parameter $\bar{\theta}_{eff}$ is well below the experimental bound $\bar{\theta}_{eff} < 10^{-10}$ and embed the four-form mechanism into a concrete stringy model - Model A, and verify its consistency with moduli stabilization in Section 5. In Section 6, we further analyze the theoretical properties and phenomenological implications of this stringy realization. Next, we discuss the constraints of swampland conjecture on strong CP problem in Section 7 and compare the results with other stringy solutions of strong CP problem in Section 8. Finally, we conclude our results in Section 9.

\section{Background}
In this section, we briefly review the background of Type IIA $T^6/(\mathbb{Z}_2 \times \mathbb{Z}_2)$ orientifold and axion. 

\subsection{Type IIA $T^6/(\mathbb{Z}_2 \times \mathbb{Z}_2)$ Orientifold}
In this section, we review the appearance of $4D$ 4-forms in Type IIA orientifold compactifications. Our focus lies on the role of Minkowski 3-form fields within the flux-induced scalar potential. We adopt the democratic formulation, which incorporates all RR $p$-form potentials for $p=1,3,5,7$ and imposes Hodge duality relations at the level of equations of motion to avoid overcounting physical degrees of freedom. The $10D$ action splits into three sectors:
\begin{equation} \label{SIIA}  
S_{IIA}=S_{RR} + S_{NS} + S_{loc},
\end{equation}
where $S_{loc}$ accounts for localized sources (D6-branes and O6-planes).

The Ramond-Ramond sector action is:
\begin{equation} \label{SRR}  
S_{RR}= -\frac{1}{4} \sum^4_{p=0} \int_{10D} G_{2p+2} \wedge \star_{10} G_{2p+2}
\end{equation}
where $G_{2p+2}= dC_{2p+1}$ are RR field strengths. 

The Neveu-Schwarz-Neveu-Schwarz sector action is:
\begin{equation} \label{SNSNS}  
S_{NSNS}= \frac{1}{2\kappa^2_{10}}  \int_{10D} e^{-2\phi}  \left(R_{10} - \frac{1}{2} H_{3} \wedge \star_{10} H_{3} \right)
\end{equation}
where $\phi$ is the dilaton, $H_3=dB_2$ is the NSNS 3-form field strength, and $R_{10}$ is the $10D$ Ricci scalar.

We compactify on a factorized $T^6/(\mathbb{Z}_2 \times \mathbb{Z}_2)$ orientifold, decomposing $10D$ coordinates as $x^M = (x^{\mu}, y^m)$, where $x^{\mu} (\mu=0,1,2,3)$ are $4D$ Minkowski coordinates and $y^m (m=4,...,9)$ are internal toroidal coordinates. All fields are expanded in a basis of harmonic forms on the internal manifold. For the relevant $p$-forms:
\begin{equation} \label{B2C3basis}  
B_2 = \sum_i b_i \omega_i, \quad C_3= \sum_I c^I_3 \alpha_I,
\end{equation}
where ${\omega_i}$ is a basis of anti-invariant 2-forms under the orientifold projection, and ${\alpha_I}$  is a basis of invariant 3-forms. The coefficients $b_i$ and $c^I_3$ become $4D$ scalar fields correspond to $4D$ scalars representing the axionic components of the complex supergravity fields $T_i$, $U_i$ and $S$. These are explicitly defined as
\begin{equation} \label{ImTi} 
\text{Im} T_i= - \int B_2 \wedge \tilde{\omega}^i = -b^i; \quad i= 1,..., h^{(1,1)}_{-}
\end{equation}
\begin{equation} \label{ImUi} 
\text{Im} U_i= \int C_3 \wedge \beta^i = c^i_3; \quad i= 1,..., h^{3}_{+}
\end{equation}
\begin{equation} \label{ImS} 
\text{Im} S= -\int C_3 \wedge \beta^0 = -c^0_3,
\end{equation}

Upon compactification and integration over the internal dimensions, one obtains a total of $2 h^{(1,1)}_{-}+2$ Minkowski 4-forms: $F^0_4$, $F^i_4$, $F^a_4$ and $F^m_4$. Specifically, there are $h^{(1,1)}_{-}$ $F^i_4$ fluxes, $h^{(1,1)}_{-}$ $F^a_4$ fluxes, one $F^0_4$ flux and one $F^m_4$ flux \cite{bielleman2015minkowski}. The resulting effective four-dimensional scalar potential for the RR sector \cite{bielleman2015minkowski},
\begin{eqnarray} \label{VRR}
\begin{split} 
V_{RR} =  & -\frac{1}{2} [-k F^0_4 \wedge \star F^0_4 +2F^0_4 \rho_0 - 4k g_{ij} \star F^i_4 \wedge F^j_4 + 2 F^i_4 \rho_i   \\
& -\frac{1}{4k} g_{ab} F^a_4 \wedge \star F^b_4 + 2 F^a_4 \rho_a + k F^m_4 \wedge \star F^m_4 ].
\end{split}
\end{eqnarray}
The Chern-Simons couplings for the 4-form field strengths are given by:
\begin{equation} \label{rho0}
  \rho_0 = e_0 + b^i e_i + \frac{1}{2} k_{ijk} q_i b^j b^k - \frac{m}{6} k_{ijk} b_i b^j b^k - h_0 c^0_3 - h_i c^i_3,
\end{equation}
\begin{equation} \label{rhoi}
  \rho_i = e_i + k_{ijk} b^j q^k - \frac{m}{2} k_{ijk} b^j b^k,
\end{equation}
\begin{equation} \label{rhoa}
  \rho_a = q_a - m b_a,
\end{equation}
\begin{equation} \label{rhom}
  \rho_m =- m.
\end{equation}
The NSNS sector contributes to the potential as
\begin{equation} \label{VNS}
  V_{NS}=\frac{1}{2} e^{-2\phi} c_{IJ} H^I_4 H^J_4,
\end{equation}
where $c_{IJ}$ is the metric on the complex structure moduli space, and the NS internal flux satisfies $\star H^I_4 =h^I$ \cite{bielleman2015minkowski}. 

Finally, the potential from localized sources takes the form:
\begin{equation} \label{Vloc1}
  V_{loc}=\sum_a \int_{\Sigma} T_a \sqrt{-g} e^{-\phi},
\end{equation}
where $T_a$ is the tension of brane and $\Sigma$ denotes its worldvolume. 

The complete perturbative superpotential in the Type IIA $T^6/(\mathbb{Z}_2 \times \mathbb{Z}_2)$ orientifold theory is given by
\begin{eqnarray} \label{Wf}
\begin{split} 
W =  & e_0 + ih_0 S + \sum^3_{i=1} [(ie_i -a_i S -b_{ii} U_i -\sum_{j\neq i} U_j)T_i- ih_i U_i]  \\
& -q_1 T_2 T_3 - q_2 T_1 T_3 - q_3 T_1 T_2 + im T_1 T_2 T_3,
\end{split}
\end{eqnarray}
where $e_0$, $e_i$, $h_0$, $h_i$, $a_i$, $b_{ij}$, $q_i$, and $m$ are flux parameters \cite{camara2005fluxes}. The Kähler potential has the following form
\begin{equation} \label{Kp} 
K= - \ln (S + S^{*}) - \sum^{3}_{i=1} \ln (U_i + U^{*}_i) - \sum^{3}_{i=1} \ln (T_i + T^{*}_i), 
\end{equation}
where, $S$ denotes the axio-dilaton, $T_i$ represents the complex structure moduli, and $U_i$ corresponds to the volume (Kähler) moduli \cite{camara2005fluxes}.

In general, we consider stacks of $N_a$ intersecting D6-branes wrapping the factorizable 3-cycle
\begin{equation} \label{Pia}
\Pi_a = (n^1_a, m^1_a) \otimes (n^2_a, m^2_a) \otimes (n^3_a, m^3_a),
\end{equation}
along with their corresponding orientifold images, which wrap the cycles $\otimes_i (n^i_a, -m^i_a)$. Here, $n^i_a$ and $m^i_a$ are the wrapping numbers of a D-brane stack “a” along the $x^i$ and $y^i$ directions, respectively, on the $i$-th two-torus. In the $\mathbb{Z}_2 \times \mathbb{Z}_2$ IIA orientifold, the cancellation conditions for Ramond-Ramond (RR) tadpoles in the presence of fluxes are given by \cite{camara2005fluxes}:
\begin{equation} \label{tcc11}
\sum_a N_a n^1_a n^2_a n^3_a + \frac{1}{2} (h_0 m + a_1 q_1 + a_2 q_2 + a_3 q_3) =16,
\end{equation}
\begin{equation} \label{tcc21}
\sum_a N_a n^1_a m^2_a m^3_a + \frac{1}{2} (m h_1 - q_1 b_{11} - q_2 b_{21} - q_3 b_{31}) =-16,
\end{equation}
\begin{equation} \label{tcc31}
\sum_a N_a m^1_a n^2_a m^3_a + \frac{1}{2} (m h_2 - q_1 b_{12} - q_2 b_{22} - q_3 b_{32}) =-16,
\end{equation}
\begin{equation} \label{tcc41}
\sum_a N_a m^1_a m^2_a n^3_a + \frac{1}{2} (m h_3 - q_1 b_{13} - q_2 b_{23} - q_3 b_{33}) =-16.
\end{equation}

For a globally consistent intersecting D6-brane toroidal model, it should satisfy several key consistency conditions: anomaly cancellation, K-theory cancellation and supersymmetry conditions.\\
\textbf{1. Anomaly cancellation}\\
RR tadpole cancellation is achieved by an appropriate choice of visible and hidden D6-branes wrapping factorizable 3-cycles \cite{camara2005fluxes}. The model should satisfy \eqref{tcc11}-\eqref{tcc41}.\\
\textbf{2. K-Theory constraints}\\
Beyond tadpole cancellation, global consistency often requires the satisfaction of discrete K-theory constraints, which prevent anomalies associated with stable stringy solitons \cite{ibanez2012string,uranga2003chiral,marchesano2007progress}, namely,
\begin{equation} \label{dKtc}
\sum_a N_a m^1_a m^2_a m^3_a \in 4 \mathbf{Z}, \quad \sum_a N_a n^1_a n^2_a m^3_a \in 4 \mathbf{Z}, \quad \text{and permutations.}
\end{equation}
\textbf{3. Supersymmetry conditions}\\
In order to preserve $\mathcal{N}=1$ supersymmetry, we should require the SUSY condition:
\begin{equation} \label{SUSYcon}
\theta_1 + \theta_2 + \theta_3 = 0 \quad \text{mod} \quad 2\pi,
\end{equation}
where $\theta_i = \tan^{-1}(\frac{m^i R_2}{n^i R_1})$ and $R_1$, $R_2$ are the two radii along two directions of every $T^2_i$. In principle, we can always choose the parameter $U_i=R^{(i)}_2/R^{(i)}_1$ to meet the SUSY condition \eqref{SUSYcon}. 

\subsection{Axion}
The QCD axion offers a natural solution to the strong CP problem, a critical fine-tuning puzzle of the Standard Model (SM). QCD dynamics permits a CP-violating topological term proportional to the effective theta parameter $\bar{\theta}_{eff}$, while neutron electric dipole moment measurements impose an extremely tight bound $\bar{\theta}_{eff} < 10^{-10}$ with no intrinsic SM mechanism to suppress this value \cite{Conlon:2006tq,dvali2022strong,burgess2024uv,choi2023implications}. The Peccei-Quinn (PQ) framework resolves this tension by promoting the static $\bar{\theta}$ to a dynamical axion scalar, a construction naturally embedded within Type IIA $T^6/(\mathbb{Z}_2 \times \mathbb{Z}_2)$ orientifold model \cite{Peccei:1977ur,Peccei:1977hh}.

The CP-odd QCD Lagrangian term reads
\begin{equation} \label{SFtildeF}
\mathcal{L} \supset \frac{1}{32\pi^2} \left( \theta - \frac{a}{f_a} \right) \text{Tr} [F_{\mu\nu} \tilde{F}^{\mu\nu}],
\end{equation}
where $F_{\mu\nu}$ denotes the $\text{SU(3)}_c$ field strength and $\bar{\theta}$ combines bare CP phases from quark mass matrices \cite{Conlon:2006tq}. The PQ mechanism introduces an anomalous global $U(1)_{\text{PQ}}$ symmetry, replacing the constant $\theta$ with a dynamical axion field $a$ via the substitution $\bar{\theta} = \theta - (a/f_a)$, where $f_a$ is the axion decay constant with mass dimension \cite{Peccei:1977ur}. 

The renormalized SM plus axion Lagrangian becomes
\begin{equation} \label{LPQ2}
\mathcal{L}= \mathcal{L}_{\text{SM}}+ \frac{1}{2} \partial_{\mu} a \partial^{\mu} a + \frac{a}{16\pi^2 f_a}  F_{\mu\nu} \tilde{F}_{\mu\nu}.
\end{equation}
QCD non-perturbative instantons explicitly break the shift symmetry $a \rightarrow a + \epsilon$, generating a periodic axion potential
\begin{equation} \label{Vins}
V_{\text{instanton}} \sim \Lambda^4_{\text{QCD}} \left(1- \cos(a/f_a) \right), 
\end{equation}
whose global minimum sits at $a=0$. This dynamically relaxes $\bar{\theta}$ to zero without fine-tuning, fully eliminating strong CP violation \cite{Conlon:2006tq}. The QCD axion mass is set by the confinement scale:
\begin{equation} \label{ma1}
m^{\text{QCD}}_a \sim \Lambda^2_{\text{QCD}}/f_a. 
\end{equation}
A necessary consistency condition for resolving the strong CP puzzle is that QCD instantons dominate the axion potential; extraneous PQ-breaking contributions shift the minimum away from $\bar{\theta}=0$ and recreate fine-tuning, known as the axion quality problem \cite{dvali2022strong,burgess2024uv,choi2023implications}. Astrophysical and cosmological observations impose a viable window for $f_a$: supernovae cooling yields $f_a > 10^9 \text{GeV}$, while stellar-mass black hole spin measurements give $f_a < 2 \times 10^{17} \text{GeV}$ \cite{Arvanitaki:2009fg}.

String compactifications universally predict an “axiverse” spectrum of light axion-like particles (ALPs originating as imaginary parts of geometric moduli) \cite{Arvanitaki:2009fg}. In our Type IIA $T^6/(\mathbb{Z}_2 \times \mathbb{Z}_2)$ orientifold model, axions correspond to $\theta_i \equiv \text{Im} T_i$, governed by the diagonal Kähler potential
\begin{equation} \label{KMA} 
K=  - \sum^{7}_{i=1} \ln (T_i + \bar{T}_i),  \quad K_{i \bar{i}} = \frac{1}{4\tau^2_i}, \quad \tau_i = \text{Re} T_i,
\end{equation}
where $T_i$ denotes the 7 moduli fields.

Kinetic terms for moduli $\tau_i$ and axions $\theta_i$ fully decouple at perturbative order \cite{Conlon:2006tq}. For example, if we identify $\theta_1$ as the QCD axion localized on the 3-cycle supporting visible-sector D6-branes carrying SM $\text{SU(3)}$ gauge interactions, canonical normalization of the axion kinetic action yields the decay constant relation in reduced Planck units $M_{pl}=1$:
\begin{equation} \label{fa1} 
f_a= \frac{1}{4\pi \tau_1 \sqrt{2}}.
\end{equation}
Taking the modulus vacuum value $\tau_1 \sim 20$ gives $f_a \sim 6.85 \times 10^{15}\text{GeV}$ ($M_{pl}=2.435 \times 10^{18}\text{GeV}$),  well within the observational band required to solve the strong CP problem \cite{Arvanitaki:2009fg,Honecker:2013mya}.

\section{Four-Form Flux Mechanism for Strong CP Problem}
In this section, we will explain how to solve the strong CP problem in Type IIA orientifold model using four-form flux mechanism. 

\subsection{The four-form flux mechanism: a brief review}
The 4-form flux framework provides a dynamical resolution to the strong CP problem by coupling the QCD axion to auxiliary 3-form gauge potentials, whose 4-form field strengths drive the effective CP-violating parameter $\bar{\theta}$ exactly to zero. Compared with the conventional Peccei–Quinn scenario, this construction naturally circumvents the axion quality problem and can be directly embedded into string compactifications \cite{dvali2022strong,burgess2024uv,choi2023implications}.

The key physical insight is that below the QCD scale, integrating out the strongly coupled gauge degrees of freedom generates an emergent, non-propagating 3-form field $C_{\mu\nu\lambda}$ with 4-form field strength $H=dC$. This field encodes the topological susceptibility of the QCD vacuum and is required for a consistent low-energy description of topological effects in QCD, analogous to auxiliary fields in quantum Hall effective field theories.

The formal derivation begins with the high-energy Lagrangian of a 2-form potential $B_{\mu\nu}$ coupled to the QCD gauge sector. The 3-form field strength is defined as 
\begin{equation} \label{G}
G=dB+S,
\end{equation}
where $S_{\mu\nu\lambda}$ is the Chern-Simons 3-form built from $A_{\mu}$. Here, $S$ satisfies $dS = \Omega$, with $\Omega$ being a gauge-invariant quantity that, for consistency, obeys $d\Omega= 0$. 

By introducing the axion $a$ as a Lagrange multiplier to enforce the Bianchi identity for $G$, and integrating out the 3-form field strength, one recovers the standard axion effective Lagrangian \eqref{LPQ2}. Below the QCD scale $\Lambda_{\text{QCD}}$, the strongly coupled vacuum gives rise to the emergent 4-form flux $H=dC$, normalized such that
\begin{equation} \label{HFF}
\frac{1}{12} \tilde{\Lambda}^2_{\text{QCD}} \epsilon^{\mu\nu\lambda\rho} H_{\mu\nu\lambda\rho}= \epsilon^{\mu\nu\lambda\rho} \langle F_{\mu\nu} F_{\lambda\rho} \rangle.
\end{equation}
Including this flux in the low-energy action yields an effective Lagrangian of the form
\begin{equation} \label{L2Ca}
\mathcal{L}_1(C,B) \supset  -\frac{1}{2}(\partial a)^2 + \frac{1}{4!} (\mu_a a -\bar{\theta} \tilde{\Lambda}^2_{\text{QCD}}) \epsilon^{\mu\nu\lambda\rho} H_{\mu\nu\lambda\rho} -\frac{1}{2 \cdot 4!} H_{\mu\nu\lambda\rho} H^{\mu\nu\lambda\rho},
\end{equation}
where the mass parameter scales as $\mu_a \sim \tilde{\Lambda}^2_{\text{QCD}}/f_a$. 

Integrating out the 4-form $H$ via its saddle-point equation produces the effective axion potential
\begin{equation} \label{Va}
V(a) = -\frac{1}{2} (\mu_a a - \bar{\theta} \tilde{\Lambda}^2_{\text{QCD}})^2 + \text{const.}
\end{equation}
Minimizing $V(a)$ with respect to the axion field enforces the vacuum condition
\begin{equation} \label{X0}
\mu_a a- \bar{\theta} \tilde{\Lambda}^2_{\text{QCD}}=0.
\end{equation}
At this minimum, the CP-violating topological term vanishes identically. The effective $\bar{\theta}_{eff}$ is dynamically relaxed to zero without fine-tuning of fundamental parameters, which constitutes the solution to the strong CP problem.

The mechanism extends naturally to theories with multiple 3-forms and multiple axions, as generically predicted by the string axiverse. For the illustrative case of two axions $a,b$ and two 4-form fluxes (with their Hodge-dual scalar invariants $X,Y$), the low-energy potential takes the form
\begin{equation} \label{Vab}   
V (a,b) = -W(X, Y) + (\mu_a a - \bar{\theta} \tilde{\Lambda}^2_{QCD})X + (\tilde{\mu}_a a + \mu_{\star} b - \eta \tilde{\Lambda}^2_{X})Y,
\end{equation}
where $W=\frac{1}{2} (X^2 + Y^2)$ for canonical kinetic terms. 

Extremizing the potential with respect to both axion fields gives
\begin{equation} \label{DVab}    
\frac{\partial V(a,b)}{\partial a} = \mu_a X + \tilde{\mu}_a Y \quad \text{and}=0 \quad \frac{\partial V(a,b)}{\partial b} = \mu_{\star} Y=0,
\end{equation}
whose unique solution is $X=Y=0$. Substituting back, both CP-violating linear combinations are cancelled at the vacuum, and the axions acquire physical masses
\begin{equation} \label{m4forma}    
m_a^2=\mu^2_a + \tilde{\mu}^2_a
\end{equation}
and 
\begin{equation} \label{m4formb}    
m^2_b= \mu^2_{*}.
\end{equation}

\subsection{Discussion on the axion quality problem}
The traditional axion quality problem fundamentally originates from the inevitable breaking of the global Peccei-Quinn (PQ) symmetry induced by quantum gravitational effects at the ultraviolet (UV) scale. In standard effective field theory frameworks, higher-order non-perturbative corrections, such as gravitational instantons and wormhole configurations, generate a series of PQ-symmetry-breaking potential terms for the QCD axion. These unwanted corrections shift the minimum of the axion potential, driving the effective vacuum angle $\bar{\theta}_{eff}$ far beyond the stringent experimental upper bound of $10^{-10}$ for resolving the strong CP problem, and they also distort the well-established correlation between the axion mass and its decay constant, thereby invalidating the conventional PQ axion solution at the low-energy regime.

The four-form flux mechanism provides a robust resolution to this long-standing dilemma by reinterpreting the PQ symmetry as a fundamental gauge symmetry rather than a fragile global symmetry \cite{dvali2022strong}. As elaborated by \cite{burgess2024uv}, the QCD axion in string compactifications is dual to the four-dimensional Kalb-Ramond two-form field, and the original global PQ shift symmetry is elevated to a local gauge symmetry of higher-form fields in the UV completion. Unlike global symmetries that are universally violated by quantum gravity, gauge symmetries are inherently protected against UV corrections, so the core symmetry underlying the axion’s light mass and CP-protective properties remains intact across all energy scales. Meanwhile, ubiquitous four-form fluxes in string vacua introduce rigorous flux quantization constraints, which drastically suppress the magnitude of residual higher-dimensional and non-perturbative corrections that would otherwise break the effective axion shift symmetry.

Further supplemented by the dual-field analysis from \cite{choi2023implications}, the four-form mechanism clarifies how topological couplings between three-form Chern-Simons fields and four-form field strengths regulate dangerous symmetry-breaking contributions. Additional three-form potentials from hidden sectors or gravitational backgrounds only trigger potential quality issues when two strict coupling conditions are simultaneously satisfied, and most string-derived higher-form fields fail to meet these criteria. Even for the few potentially problematic fields, the exponential suppression from non-perturbative gravitational effects and the collective protection from multiple coexisting axion modes in the string axiverse jointly keep the effective vacuum angle 
$\bar{\theta}_{eff}$ well within the experimental limit. This framework demonstrates that the four-form gauge structure naturally evades the axion quality problem while preserving all key features of the QCD axion as a viable solution to the strong CP problem.

\section{Embedding Four-Form Fluxes in Type IIA  $T^6/(\mathbb{Z}_2 \times \mathbb{Z}_2)$ Orientifolds}
In this section, we will discuss how to embed the four-form flux mechanism in Type IIA  $T^6/(\mathbb{Z}_2 \times \mathbb{Z}_2)$ orientifolds. In addition to the well-known Ramond–Ramond (RR) and Neveu–Schwarz (NS) fluxes, string compactifications also involve less-studied NS geometric fluxes, which emerge naturally in Scherk–Schwarz dimensional reduction schemes \cite{bielleman2015minkowski}. 

These geometric fluxes can be systematically defined on a factorized six-torus $T^6$ in the presence of O6-planes wrapping specific 3-cycles. When we impose a $\mathbb{Z}_2 \times \mathbb{Z}_2$ orbifold projection, only the diagonal moduli survive, leaving a reduced set of three Kähler moduli and four complex structure moduli (including the axio-dilaton). In such a configuration, the 12 independent geometric flux parameters $\omega^M_{NK}$, can be efficiently organized into a 3-vector $a_i$ and a $3 \times 3$-matrix $b_{ij}$ \cite{bielleman2015minkowski}. 

This parametrization leads to a modification of the 4-form field strengths, which take the following form \cite{bielleman2015minkowski}:
\begin{equation} \label{starF042}
  \star F^0_4 = \frac{1}{k} \left(e_0 + e_ib^i + \frac{1}{2} k_{ijk} q^i b^j b^k - \frac{m}{3!} k_{ijk} b^i b^j b^k -h_0 c^0_3 - h^i c^i_3 + b^i b_{ij} c^j_3 - b^i a_i c^0_3 \right),
\end{equation}
\begin{equation} \label{starFi42}
  \star F^i_4 = \frac{g^{ij}}{4k} \left(e_i + k_{ijk} b^j q^k -\frac{m}{2} k_{ijk} b^j b^k + b_{ij} c^j_3 - a_i c^0_3 \right),
\end{equation}
\begin{equation} \label{starHi4}
  \star H^i_4 = h^i - b^{ij}b_j,
\end{equation}
\begin{equation} \label{starH04}
  \star H^0_4 = h^0+b^i a_i.
\end{equation}
Here, $k_{ijk}$ denotes the triple intersection numbers, which are equal to 1 when $i$, $j$ and $k$ are all distinct and zero otherwise. 

The resulting four-dimensional scalar potential is \cite{bielleman2015minkowski}:
\begin{equation} \label{sp4D1}  
V = \frac{k}{2} |F^0_4|^2 + 2k \sum_{ij} g_{ij} F^i_4 F^j_4 + \frac{1}{8k} \sum_{ij} g_{ij} H^i_4 H^j_4 + k|H^0_4|^2 + V_{NS} + V_{lol},
\end{equation}
where $V_{NS}$ is contributions from the NS sector, and $V_{lol}$ represents the contributions from localized sources such as branes and orientifold planes.

We now establish the precise mapping between the parameters of the 4-form flux mechanism in effective field theory (EFT) and the fundamental quantities in Type IIA string theory. Following \cite{dvali2022strong,burgess2024uv,choi2023implications}, the low-energy EFT Lagrangian for the 4-form flux mechanism below the QCD scale is:
\begin{equation} \label{LEFT2}  
\mathcal{L}_{EFT} = -\frac{1}{2} (\partial a)^2 - \{\frac{1}{2 \cdot 4!} H_{\mu\nu\lambda\rho} H^{\mu\nu\lambda\rho} - \frac{1}{4!} \left(\mu_a a - \bar{\theta} \tilde{\Lambda}^2_{QCD} \right) \epsilon^{\mu\nu\lambda\rho} H_{\mu\nu\lambda\rho} \}.
\end{equation}
Integrating out the auxiliary 4-form field $H$ by solving its equation of motion yields the effective potential for the axion:
\begin{equation} \label{VEFT2}  
V_{EFT} (a) = \frac{1}{2} \left(\mu_a a - \bar{\theta} \tilde{\Lambda}^2_{QCD} \right)^2 .
\end{equation}
This quadratic potential has a unique minimum at $a =\bar{\theta} \tilde{\Lambda}^2_{QCD}/\mu_a$, which dynamically sets the effective CP-violating term to zero.

From \eqref{VRR},  the relevant part of the string theory scalar potential that corresponds to the EFT 4-form mechanism is the 
$F^a_4$ sector. For simplicity, we assume the metric simplification $\frac{1}{4k} g_{ab}=\delta_{ab}$ (valid for our factorized toroidal orientifold with equal cycle volumes), reducing this sector to:
\begin{eqnarray} \label{VRRFa4}
V^{4-form}_{\text{string}} =  -\frac{1}{2} (- F^a_4 \wedge \star F^a_4 + 2F^a_4 \rho_a)= \frac{1}{2} F^a_4 \wedge \star F^a_4 - F^a_4 \rho_a.
\end{eqnarray}

For toroidal orientifold model, the corrected potential for the $H^i_4$ sector
\begin{eqnarray} \label{VRRHi4}
V_{RR}(H^i_4) =  \frac{1}{2}  H^i_4 \wedge \star H^i_4  - (b_{ij} b^j - h_i) H^i_4.
\end{eqnarray}
We now compare the string theory potential \eqref{VRRHi4} with the EFT potential \eqref{LEFT2} term by term. For a single axion and single 4-form (the minimal case), we drop the indices $i,j$:
\begin{eqnarray} \label{VRRHi42}
V_{RR}(H^i_4) =  \frac{1}{2}  H_4 \wedge \star H_4  - (b_{11} b - h_1) H_4.
\end{eqnarray}
The EFT Lagrangian \eqref{LEFT2} can be rewritten in terms of the 4-form field strength $X\equiv \frac{1}{4!} \epsilon^{\mu\nu\lambda\rho} H_{\mu\nu\lambda\rho}$ as:
\begin{equation} \label{LEFT3}  
\mathcal{L}_{EFT} = -\frac{1}{2} (\partial a)^2 - \{\frac{1}{2} X^2 - \left(\mu_a a - \bar{\theta} \tilde{\Lambda}^2_{QCD} \right) X \}
\end{equation}
where we have used the identity $H_{\mu\nu\lambda\rho} H^{\mu\nu\lambda\rho}=24X^2$ and $\epsilon^{\mu\nu\lambda\rho} H_{\mu\nu\lambda\rho}=24X$. 

Comparing \eqref{VRRHi42} and \eqref{LEFT3} term by term, we obtain the following one-to-one correspondence:
\begin{equation} \label{axioncorresponding}
H^i_4 \leftrightarrow  X, \quad b_{ij} \leftrightarrow \mu_a, \quad b^i \leftrightarrow a, \quad h_i \leftrightarrow \bar{\theta} \tilde{\Lambda}^2_{QCD}.
\end{equation}
For the multi-axion case (STU model with 3 axions), the correspondence generalizes to the matrix form:
\begin{equation} \label{axioncorresponding2}
H^i_4 \leftrightarrow  X, \quad b_{ij} \leftrightarrow \mu_{ij}, \quad b^i \leftrightarrow a, \quad h_i \leftrightarrow \bar{\theta} \tilde{\Lambda}^2_{QCD}.
\end{equation}
The mechanism that solves the strong CP problem in this 4-form language requires the condition:
\begin{equation} \label{hi}
b_{ij} b^j = h_i.
\end{equation}
Furthermore, from \eqref{ImTi} we can know that the QCD axion in this mechanism should be the imaginary part of complex structure moduli $T_i$.

We now rigorously prove that minimizing the combined axion-4-form potential dynamically sets $\bar{\theta}=0$ and that this extremum is a stable minimum. We start from the full $4D$ effective Lagrangian including both the axion kinetic term and the 4-form potential:
\begin{equation} \label{Lkt4formt}  
\mathcal{L}= \frac{1}{2} K_{ij} \partial_{\mu} b^i \partial^{\mu} b^j + \frac{1}{2} H^i_4 \wedge \star H^i_4 + (b_{ij} b^j -h_i) H^i_4,
\end{equation}
where $K_{ij} =\frac{1}{4\tau^2_i} \delta_{ij}$ is the diagonal Kähler metric for the axions, and $\tau_i \equiv \text{Re}T_i$ are the real parts of the Kähler moduli.

We first integrate out the auxiliary 4-form fields $H^i_4$ by solving their equations of motion. Varying the action with respect to 
$H^i_4$ gives:
\begin{equation} \label{dSdHi4}  
\frac{\delta S}{\delta H^i_4} = \star H^i_4 - (b_{ij} b^j -h_i) =0,
\end{equation}
which yields the solution:
\begin{equation} \label{Hi4sol}  
 H^i_4 =  b_{ij} b^j -h_i.
\end{equation}
Substituting this solution back into the Lagrangian \eqref{Lkt4formt} gives the effective potential for the axion fields:
\begin{equation} \label{Veffbj}  
V_{eff}(b^j) \sim \sum_i (b_{ij} b^j -h_i)^2.
\end{equation}
For the minimal case of a single axion $b$ and single 4-form, this simplifies to:
\begin{equation} \label{Veffbjsbs4form}  
V_{eff}(b^j) \sim (b_{11} b -h_1)^2.
\end{equation}
To find the extremum of the potential, we take the first derivative with respect to the axion field $b$ and set it to zero:
\begin{equation} \label{dVeffdb1}  
\frac{dV_{eff}(b^j)}{db} \sim b_{11} (b_{11} b -h_1)=0.
\end{equation}
Since $b_{11} \neq 0$ (geometric flux is non-zero for a non-trivial compactification), the only solution is:
\begin{equation} \label{dVeffdb2}  
b_{11} b -h_1=0.
\end{equation}
Using the parameter correspondence from \eqref{axioncorresponding2}, we rewrite this condition as:
\begin{equation} \label{dVeffdb3}  
\mu_a a - \bar{\theta} \tilde{\Lambda}^2_{\text{QCD}}=0.
\end{equation}
This is exactly the condition required to cancel the CP-violating term in the QCD Lagrangian \eqref{X0}. The effective theta angle is given by:
\begin{equation} \label{barthetaeff1}  
\bar{\theta}_{eff} = \bar{\theta} - \frac{a}{f_a},
\end{equation}
where $f_a$ is the axion decay constant. Substituting $a= \bar{\theta} \tilde{\Lambda}_{\text{QCD}}/\mu_a$ and using the relation $\mu_a = \tilde{\Lambda}_{\text{QCD}}/f_a$ (from dimensional analysis and the axion mass formula $m_a =\mu_a$ in this mechanism), we find:
\begin{equation} \label{barthetaeff2}  
\bar{\theta}_{eff} = \bar{\theta} - \frac{\bar{\theta} \tilde{\Lambda}_{\text{QCD}}/\mu_a}{f_a} =\bar{\theta} - \bar{\theta} =0.
\end{equation}
This proves that the effective CP-violating parameter is dynamically relaxed to exactly zero at the potential minimum.

To confirm that this extremum is a stable minimum, we compute the second derivative of the effective potential:
\begin{equation} \label{d2Veffdb2}  
\frac{d^2V_{eff}(b^j)}{db^2} =b^2_{11}>0.
\end{equation}
Since the second derivative is strictly positive (as $b_{11}$ is a real, non-zero geometric flux parameter), the extremum at $\bar{\theta}_{eff}=0$ is indeed a global minimum of the potential.

This mechanism naturally evades the axion quality problem because all high-scale CP-violating contributions are absorbed into the flux parameters $h_i$ and $b_{ij}$. The potential minimum condition \eqref{dVeffdb2} is independent of the absolute values of these fluxes - it only requires their ratio to be equal to the axion VEV. Any additional CP-violating contributions from Planck-scale physics will simply shift the values of $h_i$ and $b_{ij}$, but the condition $\bar{\theta}_{eff}=0$ will still hold exactly. This is in contrast to conventional PQ axion models, where high-scale corrections can shift the potential minimum away from $\bar{\theta}_{eff}=0$. 

\subsection{Discrete Flux Solutions and Vacuum Existence}
In the $T^6/(\mathbb{Z}_2 \times \mathbb{Z}_2)$ orientifold if we take $h^{(1,1)}_{-}=3$, so the geometric flux matrix $b_{ij}$ is a 3×3 integer matrix and the NSNS flux vector $h_i$ is a 3-component integer vector. The condition for dynamical cancellation of $\bar{\theta}_{eff}$ is
\begin{equation} \label{bijbjhi}  
b_{ij} b^j =h_i, \, i=1,2,3,
\end{equation}
where $b^j \equiv \text{Im}T_i$ are the axion VEVs. For a fixed invertible $b_{ij}$ (i.e. $\det b \neq 0$), the system \eqref{bijbjhi} has a unique real solution
\begin{equation} \label{bjbjihi}  
b^j = (b^{-1})_{ji} h_i, \, i=1,2,3.
\end{equation}
Since the axion VEVs are continuous parameters, any integer pair $(b_{ij}, h_i)$ with $\det b \neq 0$ yields a valid solution. Hence the four‑form mechanism is not a fine‑tuning of continuous parameters but rather a discrete vacuum selection in the flux landscape: every choice of integer fluxes satisfying \eqref{bijbjhi} automatically gives $\bar{\theta}_{eff}=0$.

However, two additional constraints restrict the allowed $(b_{ij}, h_i)$:\\
1. \textbf{Flux quantisation and tadpole conditions.} The geometric fluxes $b_{ij}$ and the NSNS fluxes $h_i$ enter the RR tadpole cancellation conditions \eqref{tcc11}-\eqref{tcc41} and the K‑theory constraints \eqref{dKtc}. Only those integer matrices that are compatible with a globally consistent D6‑brane configuration are admissible. In practice, the tadpole equations force certain linear combinations of the fluxes to be fixed integers, reducing the number of free parameters.\\
2. \textbf{Consistency with moduli stabilisation.} The axion VEVs obtained from \eqref{bjbjihi} must coincide with the imaginary parts of the complex structure moduli that minimize the full STU superpotential \eqref{Wf}. This imposes a further algebraic relation linking the flux integers to the stabilised moduli values. For a given stabilisation scheme (e.g. the STU symmetric vacuum of Model A), only a subset of the flux choices that satisfy \eqref{bijbjhi} also reproduce the correct moduli VEVs.

Despite these restrictions, a large discrete set of solutions exists. For illustration, consider the simple case where $b_{ij}$ is diagonal, $b_{ij}= b_i \delta_{ij}$. Then \eqref{bijbjhi} reduces to $b_i b^i =h_i$ (no sum). For any non‑zero integer $b_i$, choosing $b_i=h_i/b^i$ gives a solution. A full scan of the flux parameter space is beyond the scope of this paper, but the existence of at least one consistent realization (Model A) already demonstrates that the four‑form mechanism is realizable in a discrete flux vacuum, not merely an EFT construction. Thus condition \eqref{bijbjhi} lifts the continuous tuning of $\bar{\theta}$ to a discrete selection principle in the string landscape.

\section{Stability Analysis and Explicit Realization in Model A}
In Section 5, we do stability analysis to demonstrate that the induced shift of the effective CP-violating parameter $\bar{\theta}_{eff}$ is well below the experimental bound $\bar{\theta}_{eff} < 10^{-10}$ and embed the four-form mechanism into a concrete stringy model - Model A, and verify its consistency with moduli stabilization.

\subsection{Quantitative analysis of anti-D6-brane backreaction}
In this subsection, we present a quantitative analysis to demonstrate that the induced shift of the effective CP-violating parameter $\bar{\theta}_{eff}$ is well below the experimental bound $\bar{\theta}_{eff} < 10^{-10}$. This analysis confirms the robustness of our four-form flux solution to the strong CP problem under de Sitter uplifting. For simplicity, we only consider STU set-up, i.e., $T_1=T_2=T_3=T$ and $U_1=U_2=U_3=U$.

The positive energy contribution from an anti-D6-brane wrapped on a 3-cycle of the internal manifold, described within the nilpotent superfield formalism, is given by \cite{cribiori2019uplifting,McAllister:2024lnt}:
\begin{equation} \label{VbarD6} 
V_{\overline{D6}} = \frac{\mu^4_1}{(\text{Re} \ T)^3} + \frac{\mu^4_2}{(\text{Re} \ T)^2 (\text{Re} \ S)},
\end{equation}
where $\mu_1$ and $\mu_2$ are parameters encoding the anti-D6-brane tension and flux contributions, $\text{Re} T \equiv \tau$ is the real part of the complex structure modulus and $\text{Re} S \equiv s$ is the real part of the axio-dilaton. The masses of the stabilized moduli are derived from the second derivatives of the AdS scalar potential $V_{\text{AdS}}$ at the supersymmetric minimum. For the STU model adopted in this work, the modulus mass matrix is diagonal to a good approximation, with eigenvalues \cite{kallosh2020mass}:
\begin{equation} \label{m2moduli} 
m^2_{\text{moduli}} \approx \frac{\partial^2 V_{\text{AdS}}}{\partial^2 \phi^2_i}|_{\phi_i = \phi^0_i} \sim a^2_i m^2_{3/2},
\end{equation}
where $a_i \sim \mathcal{O}(1)$ are coefficients from the non-perturbative superpotential, and $m_{3/2} =e^{K/2} |W|$ is the gravitino mass. If the gravitino mass is less than 100TeV scale, then electroweak hierarchy problem can be addressed by SUSY breaking \cite{linde2012supersymmetry}. 

The anti-D6-brane potential $V_{\overline{D6}}$ depends only on the real parts of the moduli, as it arises from the brane tension. However, since the scalar potential $V_{\text{AdS}}$ couples the real and imaginary parts of the moduli via the holomorphic superpotential $W(T,S,U)$, a small shift in the real part $\delta \tau$ will induce a corresponding shift in the imaginary part (axion field) $\delta b$. 

We perform a linear perturbation analysis around the AdS minimum $(\tau_0, b_0)$ where $\partial V_{\text{AdS}}/ \partial \tau = \partial V_{\text{AdS}}/ \partial b =0$. The total potential with anti-D6-brane uplifting is $V_{\text{total}}=V_{\text{AdS}} + V_{\overline{D6}}(\tau)$.  Expanding $V_{\text{tot}}$ to first order in the shifts $\delta \tau$ and $\delta b$ and requiring the new minimum to satisfy $\partial V_{\text{tot}}/ \partial \tau = \partial V_{\text{tot}}/ \partial b =0$ gives the linear system
\begin{equation} \label{newminimum1} 
V_{\tau \tau} \delta \tau + V_{\tau b} \delta b = - \partial_{\tau} V_{\overline{D6}}, 
\end{equation}
\begin{equation} \label{newminimum2} 
V_{\tau b} \delta \tau + V_{b b} \delta b = 0, 
\end{equation}
where we have defined
\begin{equation} \label{VtautauVtaubVbb} 
V_{\tau \tau} \equiv \frac{\partial^2 V_{\text{AdS}}}{\partial \tau^2}|_0, \quad V_{\tau b} \equiv \frac{\partial^2 V_{\text{AdS}}}{\partial \tau \partial b}|_0, \quad V_{bb} \equiv \frac{\partial^2 V_{\text{AdS}}}{\partial b^2}|_0.
\end{equation}
Solving \eqref{newminimum1} and \eqref{newminimum2} yields:
\begin{equation} \label{deltab} 
\delta b= \frac{V_{\tau b}/ V_{bb}}{V_{\tau \tau} - V^2_{\tau b}/V_{bb}} \partial_{\tau} V_{\overline{D6}} \equiv \frac{V_{\tau b}}{V_{bb}} \frac{ \partial_{\tau} V_{\overline{D6}} }{m^2_{\tau}},
\end{equation}
where the physical mass eigenvalue of the Kähler modulus is
\begin{equation} \label{m2tau} 
m^2_{\tau} \equiv V_{\tau \tau} - \frac{V^2_{\tau b}}{V_{bb}}.
\end{equation}
From the parameter correspondence established in Section 4, the effective theta angle is given by:
\begin{equation} \label{thetabar} 
\bar{\theta} = \frac{h_i - b_{ij} b^j}{\tilde{\Lambda}^2_{\text{QCD}}}.
\end{equation}
A shift in the axion field $\delta b^j$ therefore induces a shift in
\begin{equation} \label{deltathetabar} 
\delta \bar{\theta} = -\frac{ b_{ij} \delta b^j}{\tilde{\Lambda}^2_{\text{QCD}}}.
\end{equation}
For the minimal case of a single axion and single geometric flux $b_{11} \sim \mathcal{O}(1)$, this simplifies to:
\begin{equation} \label{absdeltathetabar1} 
|\delta \bar{\theta}| \sim \frac{|\delta b|}{\tilde{\Lambda}^2_{\text{QCD}}}.
\end{equation}
Substituting the expression for $\delta b$ and using $\partial V_{\overline{D6}}/\partial \tau \sim 3 V_{\overline{D6}} / \tau$ from \eqref{VbarD6}, we obtain:
\begin{equation} \label{absdeltathetabar2} 
|\delta \bar{\theta}| \sim \frac{b_{ij}}{\tilde{\Lambda}^2_{\text{QCD}}} \cdot \frac{V_{\overline{D6}}}{\tau m^2_{\tau}}
\end{equation}
We now impose the experimental requirement $|\delta \theta| < 10^{-11}$ (one order of magnitude below the neutron EDM bound to account for theoretical uncertainties). Using the following values as an example: 
\begin{itemize} 
\item $b_{ij} \sim \mathcal{O}(1)$ (geometric flux parameter)
\item $\tilde{\Lambda}_{\text{QCD}} \sim 200\text{MeV} \sim 10^{-19}M_{pl}$
\item $\tau \approx 50$ 
\item $m_{\tau} \sim 1\text{TeV} \sim 10^{-16}M_{pl}$
\end{itemize}
we find the sufficient condition: 
\begin{equation} \label{VbarD6sufficientcondition} 
V_{\overline{D6}} < 5 \times 10^{-80} M^4_{pl}.
\end{equation}
The observed cosmological constant $\Lambda \approx 10^{-120} M^4_{pl}$ is about 40 orders of magnitude smaller than this upper bound, demonstrating that the backreaction-induced shift of $\bar{\theta}$ is completely negligible for all physically relevant values of the anti-D6-brane energy density.

\subsection{Explicit Realization in Model A}
In this subsection, we demonstrate how the key ingredients of the four-form flux mechanism are naturally realized within the specific context of Model A, which is a globally consistent $T^6/(\mathbb{Z}_2 \times \mathbb{Z}_2)$ Type IIA intersecting D6-brane orientifold model \cite{camara2005fluxes}. More details of Model A can be found in Appendix.

In Model A we adopt the standard three-stage STU and KL moduli stabilization scheme, and we verify full compatibility with the embedded four-form flux mechanism at every stage:\\
\begin{itemize} 
\item Supersymmetric Minkowski baseline vacuum: The superpotential
\begin{equation} \label{SWSTU}
W = -\tilde{a}TS - \tilde{b}TU + e_0 + ih_0 S - ih_1 U - i \tilde{e}T + W_{np},
\end{equation}
where we have taken $T_1=T_2=T_3=T$ and $U_1=U_2=U_3=U$ and F-term conditions $W=\partial_{T} W= \partial_{S} W = \partial_{U} W=0$ fix all moduli VEVs $t_0$, $s_0$, $u_0$ without flat directions. The geometric flux matrix $b_{ij}$ and NSNS fluxes $h_i$ entering the four-form potential appear as constant background parameters in the superpotential, not dynamical moduli, so they do not disturb the F-term minimization conditions.
\item AdS vacuum via small $\Delta W$ perturbation: A suppressed flux term $\Delta W =f_0 T^3$ generates TeV-scale gravitino mass 
$m_{3/2}$. The induced AdS potential $V_{\text{AdS}}=-3 m^2_{3/2}$ only depends on modulus real parts, decoupled from the axion potential controlled by imaginary moduli $b^j$; no cross-term shifts the $\bar{\theta}_{eff}=0$ vacuum condition. 
\item de Sitter uplifting via anti-D6-branes: The positive uplifting energy $V_{\overline{D6}}$ only couples to 
$\tau=\text{Re}T$, dilaton $s=\text{Re}S$, and leaves the axionic imaginary components $b^j$ as protected shift-symmetric scalars at leading order.
\end{itemize}
All global consistency criteria of Model A remain intact after four-form embedding: RR tadpole cancellation equations, K-theory 
$\mathbb{Z}_2$ discrete constraints, and $N=1$ SUSY wrapping-angle condition $\theta_1 + \theta_2 + \theta_3 = 0$ mod $2\pi$
are still identically satisfied with the chosen D6-brane stacks in Table 1 in Appendix. No additional flux fine-tuning is introduced to accommodate the strong CP solution.
The uplifting potential from wrapped anti-D6-branes in nilpotent superfield formalism reads
\begin{equation} \label{VbarD62} 
V_{\overline{D6}} = \frac{\mu^4_1}{(\text{Re} \ T)^3} + \frac{\mu^4_2}{(\text{Re} \ T)^2 (\text{Re} \ S)}.
\end{equation}
All stabilized moduli masses from AdS potential second derivatives scale with the gravitino mass: $m_{\text{mod}} \sim m_{3/2} \sim \text{TeV}$. 

We perform linear perturbation around the stabilized vacuum $(\tau_0, b_0)$ to compute the axion shift $\delta b$ induced by $V_{\overline{D6}}$. The total potential $V_{\text{tot}} = V_{\text{AdS}} + V_{\overline{D6}}$ yields coupled linearized extremum equations for real modulus shift $\delta \tau$ and axion shift $\delta b$:
\begin{equation} \label{extreeq1} 
m^2_{\tau} \delta \tau +  V_{\tau b} \delta b = -\partial_{\tau} V_{\overline{D6}},
\end{equation}
\begin{equation} \label{extreeq2} 
V_{\tau b} \delta \tau + m^2_{b} \delta b = 0,
\end{equation}
where we have used the fact that in STU set-up the off-diagonal terms are much smaller than the diagonal terms, i.e., $V_{\tau b} \ll m^2_{\tau} m^2_b$. Therefore, from \eqref{m2tau} we have $m^2_{\tau} \approx V_{\tau \tau}$ and we have defined $m^2_b \equiv V_{bb}$. Solving for the axion perturbation gives
\begin{equation} \label{deltab2} 
\delta b \sim \frac{V_{\tau b}}{m^2_b} \frac{ \partial_{\tau} V_{\overline{D6}} }{m^2_{\tau}}.
\end{equation}
The corresponding shift of effective theta is
\begin{equation} \label{deltaefftheta} 
\delta \bar{\theta} = - \frac{b_{ij} \delta b^j}{\tilde{\Lambda}^2_{\text{QCD}}}.
\end{equation}
Imposing the hierarchy condition $V_{\overline{D6}} \ll m^2_{\text{mod}}$ guarantees negligible backreaction.

\section{Phenomenological Checks}
We derive the phenomenologically accessible signatures of the QCD axion in the set-up we discussed in this paper, focusing on the axion decay constant window, the axion-photon coupling strength and the detection reach of experiments operating over the next 5–10 years. All predictions are derived consistently from the four-form flux embedded string construction and the STU moduli stabilization framework.

\subsection{Predicted Range of the QCD Axion Decay Constant}
In this set-up the QCD axion corresponds to the imaginary part of the complex structure modulus associated with the 3-cycle supporting the visible QCD brane stack. The decay constant is related to the real part of the modulus $\tau= \text{Re} T$ via
\begin{equation} \label{fa2} 
f_a= \frac{M_{pl}}{4\pi \tau_1 \sqrt{2}},
\end{equation}
where $M_{pl} = 2.435 \times 10^{18}\text{GeV}$ is the reduced Planck mass. We take the range of $\tau_1 \in [50,500]$ as an example. This translates to a predicted QCD axion decay constant window:
\begin{equation} \label{farange} 
2.74 \times 10^{14} \text{GeV} \leq f_a \leq 2.74 \times 10^{15} \text{GeV}.
\end{equation}
This range lies fully within the established astrophysical bounds $10^9\text{GeV} \leq f_a \leq 2 \times 10^{16} \text{GeV}$ from supernova cooling and stellar black hole spin measurements, occupying the so-called intermediate axion window between low-scale hadronic axions and conventional GUT-scale string axions.

The QCD axion mass is determined by the standard relation $m_a = \Lambda^2_{\text{QCD}}/f_a$. Using $\Lambda_{\text{QCD}} \approx 200 \text{MeV}$ together with the decay constant window \eqref{farange} gives $1.5 \times 10^{-8} \text{eV} \leq m_a \leq 1.5 \times 10^{-7}\text{eV}$. This mass range falls squarely within the design sensitivity of next-generation ultralight axion dark matter detectors. In particular, DM Radio and ABRACADABRA‑1m will be able to probe the entire predicted parameter space \cite{hook2018radio,ouellet2019first}. A non‑observation by these experiments would therefore rule out the QCD axion in our construction, providing a falsifiable test of the string‑theoretic strong CP solution.

\subsection{Axion-Photon Coupling}
The axion-photon interaction, which underpins most experimental axion searches, arises from the chiral anomaly of the QCD axion. The effective Lagrangian takes the form
\begin{equation} \label{axionphotonL}
\mathcal{L} \supset \frac{g_{\alpha \gamma}}{4} a F_{\mu\nu} \tilde{F}^{\mu\nu},
\end{equation}
where the coupling strength is parametrized as $g_{\alpha \gamma} = \frac{\alpha}{\pi f_a} \mathcal{C}_{\gamma}$. Here $\alpha \approx \frac{1}{137}$ is the fine-structure constant, and $\mathcal{C}_{\gamma}$ is an order-one model-dependent coefficient encoding the ratio of electromagnetic to color anomalies $E/N$. $\mathcal{C}_{\gamma}$ is typically $\mathcal{O}(1)$, such as $C_{\gamma} \sim 1.2-1.5$. Substituting the predicted $f_a$ window yields the coupling range:
\begin{equation} \label{gagammaarange} 
1 \times 10^{-18} \text{GeV}^{-1} \leq |g_{a \gamma}| \leq 5 \times 10^{-18} \text{GeV}^{-1}.
\end{equation}
This coupling is suppressed relative to low-scale axion models, but remains testable via both cosmological observations and next-generation ultralow-mass dark matter detectors \cite{di2020landscape}.

\subsection{Comparison with Near-Future Experimental Sensitivities}
We assess the detectability of the Model A QCD axion against the projected reach of leading axion search facilities, as well as complementary cosmological probes:
\begin{itemize}
 \item Microwave and high-frequency cavity experiments (ADMX-HF): The ADMX-HF program extends conventional resonant cavity searches to higher axion masses $\sim 1-100 \mu\text{eV}$, corresponding to decay constants $f_a \sim 10^{12}-10^{13} \text{GeV}$. This parameter range lies below the decay constant window predicted by Section 6.1, so the QCD axion in our construction is not accessible to ADMX-HF \cite{asztalos2010squid,braine2020extended}.
 \item Light-shining-through-wall experiments (ALPS II): The ALPS II experiment at DESY probes axion-photon couplings down to 
$\sim 2 \times 10^{-11} \text{GeV}^{-1}$, corresponding to $f_a \sim 10^{10} \text{GeV}$. 
ALPS II is sensitive to more strongly coupled axion-like particles that may arise from other moduli in the compactification \cite{bahre2013any}.
\item Solar axion telescopes (IAXO): The International Axion Observatory (IAXO) and its precursor BabyIAXO aim for a sensitivity of $g_{\alpha \gamma} \sim 10^{-12} \text{GeV}^{-1}$ for solar axions, corresponding to decay constants around $10^{11} \text{GeV}$, so the QCD axion in our construction is not accessible to IAXO \cite{armengaud2014conceptual,irastorza2013international}.
\item Ultralow-mass axion dark matter detectors: Next-generation low-frequency search programs, including DM Radio, ABRACADABRA-1m, and superconducting qubit-based axion detectors, are designed to probe axion masses in the $10–100 \text{neV}$ range, which directly maps to $f_a \sim 10^{14} - 10^{15} \text{GeV}$. With projected sensitivities reaching $g_{\alpha \gamma} \sim 10^{-18} \text{GeV}^{-1}$ after 5–10 years of operation, these experiments will fully cover the QCD axion parameter space predicted by this set-up and provide a direct test of our string construction \cite{hook2018radio,ouellet2019first}.
\item The axion-photon coupling induces a rotation of the CMB polarization plane during reionization, with a predicted angle $\Delta \beta < 0.1^{\text{o}}$ \cite{fedderke2019axion}. The current Planck limit $\Delta \beta < 0.27^{\text{o}}$ is consistent with this prediction, while next-generation missions such as CMB-S4 and LiteBIRD are projected to achieve $\sim 0.01^{\text{o}}$ sensitivity in the coming decade. This offers a complementary cosmological probe that is independent of ground-based direct detection \cite{ade2022bicep}.
\end{itemize}
A key distinguishing feature of our four-form flux mechanism is that the axion quality problem is fully resolved: the vanishing of $\bar{\theta}_{eff}$ remains exact even at the high decay constant scale, with no residual CP shift induced by high-scale UV corrections. This sets our scenario apart from conventional Peccei–Quinn axion models, where large $f_a$ generically introduces a fine-tuned residual $\bar{\theta}$. A future detection of an axion in this intermediate window, combined with improved neutron electric dipole moment constraints at the $10^{-11}$ level, would therefore provide strong evidence for the string-theoretic four-form flux solution to the strong CP problem. Using the axion-photon coupling from \eqref{gagammaarange}, $|g_{a \gamma} | \in [1 \times 10^{-18}, 5 \times 10^{-18}] \text{GeV}^{-1}$, and the axion mass $m_a= \Lambda^2_{\text{QCD}}/f_a$ with $\Lambda_{\text{QCD}} \approx 200 \text{MeV}$ giving $m_a \in [1.5 \times 10^{-8}, 1.5 \times 10^{-7}] \text{eV}$, the QCD axion of our set-up occupies the parameter space region shown in Figure 1. This region lies within the target sensitivity of next-generation DM Radio/ABRACADABRA searches.
\begin{figure}[h]
\centering
\includegraphics[width=0.85
\linewidth]{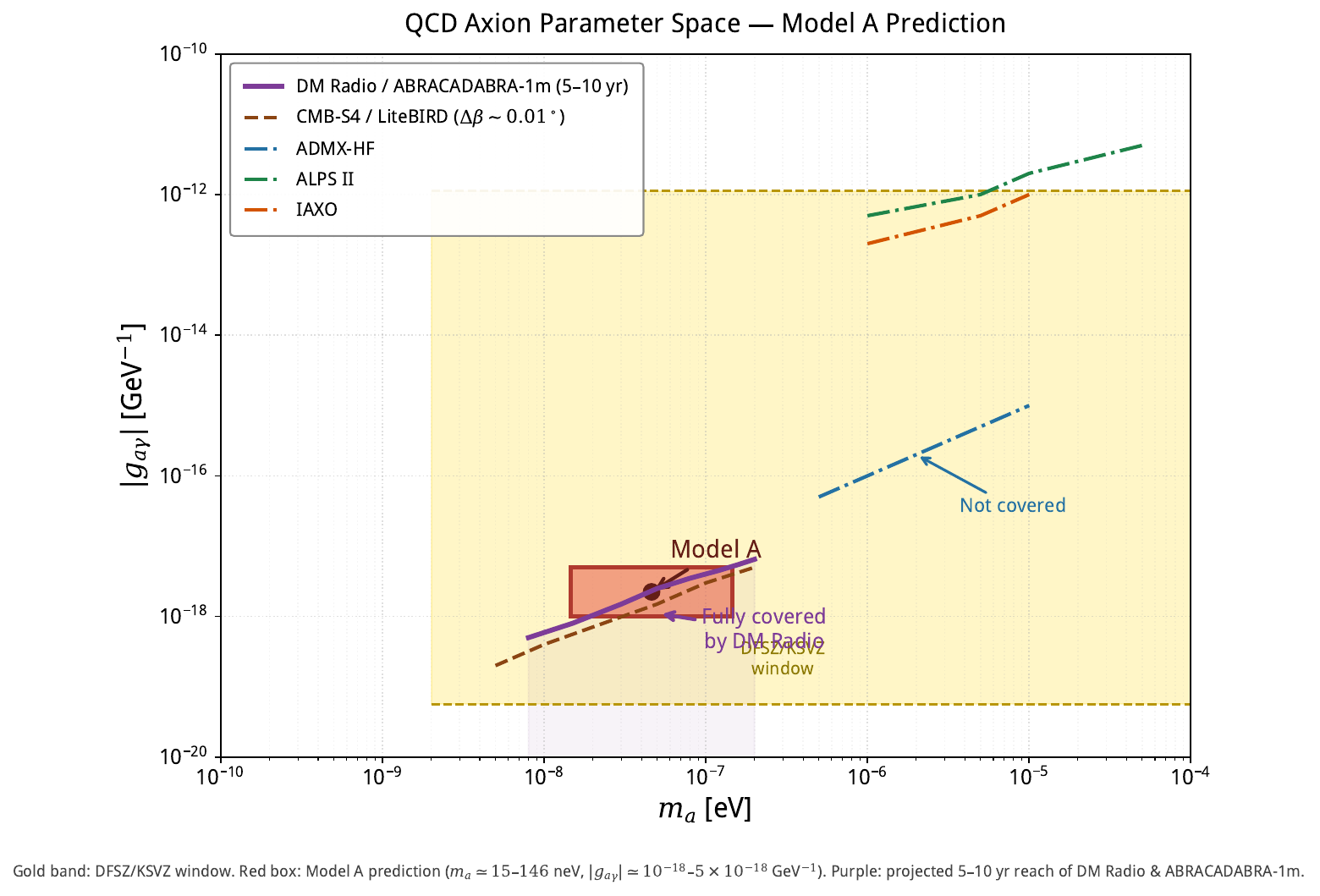}
\caption{QCD axion parameter space: axion mass $m_a$ vs.\ axion
--photon coupling $|g_{a\gamma}|$. The gold band shows the conventional QCD axion window (DFSZ--KSVZ). The red box is the Model A prediction derived from the four-form flux mechanism in this work. The purple curve indicates the target sensitivity of DM~Radio and ABRACADABRA-1m, which fully covers our predicted region. Other experimental sensitivities (ADMX-HF, ALPS~II, IAXO) and the CMB polarisation bounds (CMB-S4/LiteBIRD) are also shown \cite{asztalos2010squid,braine2020extended,bahre2013any,armengaud2014conceptual,irastorza2013international,hook2018radio,ouellet2019first}.}
\label{fig:axion_space}
\end{figure}

\section{Swampland Conjecture and Strong CP problem}

\subsection{General Considerations}
The strong CP problem has two standard resolutions within effective field theory (EFT): the Peccei–Quinn (PQ) mechanism with an axion \cite{Peccei:1977hh,weinberg1978new,wilczek1978problem}, and explicit BSM sources that drive $\theta_{\text{QCD}}$ small without a light pseudoscalar. From the swampland perspective \cite{vafa2005string,ooguri2007geometry}, neither resolution is automatically safe, because both rely on finely tuned or long-distance features that may be in tension with conjectural consistency requirements on any EFT that admits a UV completion in quantum gravity. Three swampland conjectures are relevant:

\begin{itemize}
\item Weak gravity conjecture (WGC) for axions (aWGC): For a periodic scalar $\phi$ with decay constant $f_{\phi}$ and a charged particle of charge $q$ under the associated $U(1)$, the WGC demands a superextremal state with $m<q f_{\phi}$ (in appropriate units) so that the EFT cannot be arbitrarily extrapolated \cite{arkani2006string}. In the QCD axion context this implies a lower bound $f_a \leq M_{pl}$ (more precisely, there must exist some instanton-coupled particle with $m \leq q f_{\phi}$), but also - when combined with the magnetic version \cite{harlow2016wormholes} - suggests that an axion with $f_a \gg M_{pl}$ is swampland. The PQ scale favoured by astrophysical and laboratory bounds sits comfortably below $M_{pl}$, so the QCD axion is not ruled out. However, the quality problem - why higher-dimension PQ-breaking operators are absent or sufficiently suppressed to keep $\bar{\theta}_{eff}< 10^{-10}$ - becomes sharper: if one tries to explain the suppression by a symmetry that looks accidental from the UV point of view, the WGC and completeness often forces additional light states that spoil the quality \cite{kamionkowski1992planck,holman1992solutions}.
\item Swampland distance conjecture (SDC): As a scalar $\phi$ approaches infinite distance in moduli space, an infinite tower of states becomes light with mass scaling $m \sim M_{pl} e^{-\alpha \Delta \phi/M_{pl}}$ \cite{ooguri2007geometry}. If one attempts to dial $\theta_{\text{QCD}} \rightarrow 0$ by adjusting a modulus (e.g. a complex structure or dilation that controls $\bar{\theta} = \theta + \text{arg} \text{det} M_q$), the distance conjecture suggests that the “small $\theta$" region is at best a finite-distance corner, and that a truly dynamical relaxation to $\bar{\theta}_{eff}=0$ via rolling a very light field may be cut off by tower emergence. This does not rule out the axion mechanism but constrains the way $\bar{\theta}_{eff}$ evolves: a landscape of $\bar{\theta}_{eff}$ values formed by different vacuum states is preferable to continuously tuning it to an arbitrarily small value within a single effective field theory (EFT) framework without encountering new physics.
\item AdS moduli stability/dS conjectures: If one embeds the QCD axion in a compactification that produces a (meta)stable dS or a controlled AdS, the requirement that all moduli be stabilized with masses above the Hubble scale (or obeying the refined dS conjecture \cite{obied2018sitter}) limits how many independent axion-like directions can remain light. In particular, if the PQ symmetry is taken as an accidental low-energy remnant of a high-scale symmetry broken by stringy instantons, the same instantons that generate the axion potential also generate $\theta$-dependent terms; matching $\bar{\theta}_{eff}<10^{-10}$ then requires either $f_a$ in the usual window and a vacuum misalignment $\langle a \rangle /f_a$ tuned to less than $10^{-10}$ (unless $m_u=0$ or similar infrared accident is invoked). The “swampland” perspective views such fine-tuning with skepticism, unless there exists some underlying shift symmetry - broken only by QCD instanton effects - that enables a string-theoretic construction.
\end{itemize}

\subsection{The Constraints of Swampland Conjecture on Type IIA Intersecting D6-brane models}
We now specialize the above swampland bounds to the concrete UV-complete string construction studied in this paper.  
The QCD axion is identified as the imaginary part of the complex structure modulus $T$, with canonical normalization yielding the decay constant relation
\begin{equation} \label{fatau1}
f_a =\frac{M_{pl}}{4\pi \tau \sqrt{2}}, \quad \tau \equiv \text{Re} T.
\end{equation}
We analyze swampland restrictions unique to this intersecting D6-brane setup, combining quantum gravity consistency criteria with the model’s global stringy constraints (RR tadpole cancellation, K-theory discrete anomalies, Freed–Witten consistency conditions etc.).

\subsubsection{Axion Decay Constant Window from aWGC and SDC Compatibility}
Two swampland-forbidden parameter regimes emerge from scanning moduli space:
\begin{itemize}
\item Small $\tau$: $\tau \ll 50$ produces $f_a \gg 2.74 \times 10^{15} \text{GeV}$, satisfying the axionic WGC. D2-brane instantons become light and generate dominant CP-violating potential contributions that overwhelms QCD instanton dynamics, moving the vacuum away from $\bar{\theta}=0$.
\item Excessively large $\tau$: $\tau \gg 500$ drives the complex structure modulus toward the moduli-space asymptotic boundary, triggering the SDC infinite light-state tower. The low-energy axion EFT breaks down, and multi-axion kinetic mixing between the QCD axion and spectator moduli axions ($\text{Im}S$, $\text{Im}U_i$) corrupts the CP-conserving minimum.
\end{itemize}
Additionally, the superpotential perturbative correction $\Delta W \propto T^3$ imposes a tight bound on the flux parameter $f_0$ (denoted $m$ in the full STU superpotential). 

\subsubsection{Evasion of the Axion Quality Problem via Higher-Form Gauge Symmetries}
Model A resolves the swampland no-global-symmetry obstruction to PQ axions, because all axionic shift symmetries descend from $10D$ RR $C_3$ and NSNS $B_2$ $p$-form gauge transformations, not fundamental global symmetries. The axions $\text{Im}T_i$, $\text{Im}S$, $\text{Im}U_i$ are zero-form components of compactified higher-form potentials, and their shift invariance is a residual gauge symmetry of the full $10D$ theory, immune to pure Planck-scale global-symmetry breaking operators.

Critical swampland tension persists, however, from background fluxes $b_i$, $b_{ij}$, $h_i$ intrinsic to the orientifold compactification, which explicitly break the axion shift symmetry and introduce flux-induced CP-violating potential terms. For the four-form flux strong CP mechanism to remain landscape-compatible, a flux hierarchy must hold: $V_{\text{flux}} \ll \Lambda^4_{\text{QCD}}$. This translates to a linear consistency relation on internal fluxes derived from the 4-form–axion duality dictionary: $b_{ij} b^j =h_i$. Violating this flux relation generates a dominant flux potential that shifts the axion vacuum to $\bar{\theta}=0$, rendering the strong CP solution invalid and placing the flux vacuum in the Swampland. The RR tadpole cancellation conditions and discrete $\mathbb{Z}_2$ K-theory anomaly constraints of the $T^6/(\mathbb{Z}_2 \times \mathbb{Z}_2)$ orientifold further restrict the integer flux sets $(b_i, b_{ij}, h_i)$ that can satisfy this swampland flux hierarchy simultaneously. Any flux configuration failing both tadpole cancellation and the $b_{ij} b^j =h_i$ relation is UV-incomplete and swampland-excluded.

\subsubsection{de Sitter Swampland Restrictions on Anti-D6 Uplift}
Model A constructs its metastable dS vacuum via uplifting an AdS STU minimum using the positive tension contribution of anti-D6-branes, whose potential is \eqref{VbarD6}. The dS swampland gradient bound enforces tight limits on the uplift mass scales 
$\mu_{1,2}$. Two landscape requirements must coexist:\\
1. The uplift energy $V_{\overline{D6}}$ must be a small perturbation on the negative AdS vacuum energy $V_{\text{AdS}}=-3m^2_{3/2}$, such that $V_{\overline{D6}} \ll m^2_{\text{mod}}$ to suppress anti-D6 backreaction on stabilized moduli. Strong backreaction shifts moduli VEVs away from the values required for $b_{ij} b^j =h_i$, breaking the four-form flux strong CP mechanism.\\
2. The full scalar potential must satisfy $|\nabla V|/V \geq \frac{c}{M_{pl}}$ globally. Tuning $\mu_{1,2}$ to violate this gradient condition generates unconstrained dS vacua that cannot be UV-completed within controlled Type IIA flux compactifications, hence belonging to the Swampland.

\subsection{Combined Consistency Summary for Landscape Strong CP Vacua}
For a $T^6/(\mathbb{Z}_2 \times \mathbb{Z}_2)$ intersecting D6-brane vacuum like Model A to simultaneously resolve the strong CP problem and evade all swampland restrictions, six joint criteria must hold:\\
1. The QCD axion decay constant lies in $10^9 \text{GeV} < f_a < 2 \times 10^{16}\text{GeV}$, satisfying both aWGC and SDC.\\
2. Internal geometric and NSNS fluxes obey $b_{ij} b^j =h_i$, suppressing flux-induced CP-violating axion potentials below $\Lambda^4_{\text{QCD}}$.\\
3. All RR tadpole cancellation and K-theory discrete anomaly conditions are satisfied for the D6-brane wrapping numbers and flux quanta.\\
4. The STU superpotential non-perturbative correction takes the form $\Delta W \propto T^3$, with flux parameter 
$|f_0| \ll 1$ to produce TeV-scale $m_{3/2}$.\\
5. Anti-D6 uplift parameters $\mu_{1,2}$ satisfy the dS swampland gradient bound and minimal backreaction hierarchy 
$V_{\overline{D6}} \ll m^2_{\text{mod}}$.

We can expect that only narrow slices of the full flux-moduli parameter space satisfy all six rules simultaneously. All remaining parameter combinations either fail to dynamically relax $\bar{\theta} \rightarrow 0$ or violate fundamental quantum-gravity swampland consistency conjectures, making them UV-incomplete effective theories excluded from the string Landscape.

\subsection{Concrete Numerical Example}
To illustrate that the six joint criteria of Section 7.3 select a narrow but non‑empty region of parameter space, we present an explicit benchmark point in Model A (with STU symmetry). All fluxes and moduli VEVs are chosen to simultaneously satisfy the six conditions listed in Section 7.3.

We take the following the quantities:
\begin{itemize} 
\item Geometric fluxes $b_{ij}$: $b_{11}=b_{22}=b_{33}=1$ and other fluxes are zero; 
\item NSNS fluxes $h_i$: $h_1 =h_2 =h_3=1$;
\item Axion VEVs $b^j$: $b^1=b^2=b^3=1$; 
\item Kähler modulus real part $\tau \equiv \text{Re}T$: 50;
\item Flux parameter $f_0$ (coefficient of $\Delta W \propto T^3$): $1.5 \times 10^{-8}$;
\item Anti‑D6 uplift parameters $\mu_{1,2}$: $10^{-10}M_{pl}$;
\item D6‑brane wrapping numbers: As in Table 1 (Model A).
\end{itemize}
With this choice, the six criteria are all met:
\begin{itemize} 
\item $f_a$ window: $f_a = 2.74 \times 10^{15} \text{GeV}$ lies inside $[10^9, 2 \times 10^{16}]\text{GeV}$; 
\item Flux condition: $b_{ij}b^j=h_i$ holds exactly;
\item Global consistency: Tadpole and K‑theory conditions are satisfied by the Model A brane configuration.; 
\item Gravitino mass: $m_{3/2} \sim \text{TeV}$;
\item Backreaction hierarchy: $V_{\overline{D6}} \ll m^2_{\text{mod}}$;
\item dS swampland gradient bound: For the chosen $\mu_{1,2}$ one verifies: $|\nabla V|/V \approx c/M_{pl}$, where $c \sim \mathcal{O}(1)$.
\end{itemize}

We examine a 10 percent change to the benchmark parameters while keeping all other quantities fixed:\\
\textbf{Increase the axion VEV $b^1$ by 10\%} ($b^1 \rightarrow 1.1$):\\
The flux condition becomes $b_{1j}b^j=1.1 \neq h_1 =1$. Consequently, the effective theta angle no longer vanishes: $\bar{\theta}_{eff} \neq 0$. Criterion 2 (flux condition) is violated. The strong CP problem is not solved, and the vacuum fails to realize the four‑form mechanism.

This example demonstrates that the parameter space of the strong CP solution is not a generic feature of the flux landscape but requires a specific, tuned combination of fluxes and moduli.

\section{Comparison with Other Stringy Strong CP Solutions}
To emphasize the significance of our Type IIA intersecting D6-brane four-form flux mechanism, we present a comparison with two widely studied string-theoretic axion solutions to the strong CP problem: Type IIB Large Volume Scenario (LVS) axions and heterotic string axions \cite{Conlon:2006tq,ibanez2012string,dvali2022strong,burgess2024uv,choi2023implications,kallosh2020mass,kachru2003sitter,buchbinder2015heterotic}.

\textbf{Detailed Comparison of Key Criteria:}
\begin{itemize} 
\item Global Consistency: An advantage of our construction is the existence of an explicit, fully verified globally consistent brane configuration. In contrast, Type IIB LVS models rely on hypothetical strongly warped regions whose global properties are poorly understood, while heterotic string models struggle with simultaneous gauge anomaly cancellation and moduli stabilization.
\item Axion Quality Problem: Our four-form flux mechanism is the only one among the three that provides a complete, UV-complete resolution of the axion quality problem. Both IIB LVS and heterotic axion models are vulnerable to high-scale CP-violating corrections that can reintroduce a non-zero $\bar{\theta}_{eff}$. 
\item Experimental Testability: The axion decay constant in our model falls precisely in the range for experimental detection. The predicted CMB polarization rotation signal is within the sensitivity of upcoming experiments, making our scenario falsifiable in the next decade. In contrast, the higher decay constants in IIB LVS and heterotic models make their experimental verification extremely challenging.
\end{itemize}

Our Type IIA intersecting D6-brane four-form flux mechanism represents a compelling solution to the strong CP problem within string theory. It combines full theoretical consistency (tadpole cancellation, K-theory constraints etc.) with exceptional phenomenological viability: it completely resolves the axion quality problem, predicts an axion decay constant in the optimal experimental window, and yields a testable CMB polarization signal. Unlike alternative string axion scenarios, which either suffer from unresolved theoretical inconsistencies or are effectively unobservable, our construction provides a concrete, falsifiable framework that bridges fundamental string theory and experimental particle physics and cosmology \cite{Conlon:2006tq,ibanez2012string,dvali2022strong,burgess2024uv,choi2023implications,kallosh2020mass,kachru2003sitter,buchbinder2015heterotic}.

\section{Conclusions}
In this work, we present a string-theoretic realization of the four-form flux solution to the strong CP problem, explicitly embedded in Type IIA toroidal orientifold with intersecting D6-branes.

Starting from the democratic formulation of ten-dimensional Type IIA supergravity, we perform a dimensional reduction to derive the full set of four-dimensional four-form field strengths from RR, NSNS and geometric fluxes. We establish a precise one-to-one correspondence between the parameters of the effective field theory four-form mechanism and the fundamental ingredients of the string compactification, including geometric flux coefficients, NSNS/RR flux quanta, and the axionic components of moduli. We  prove that minimizing the combined axion-four-form scalar potential dynamically relaxes the effective QCD CP-violating parameter $\bar{\theta}_\text{eff}$ to exactly zero, and confirm via second-derivative analysis that this vacuum is a stable global minimum. A key virtue of this construction is that it naturally evades the axion quality problem: all high-scale CP-violating ultraviolet corrections are absorbed into quantized background fluxes, and the condition $\bar{\theta}_\text{eff}=0$ holds exactly regardless of Planck-scale contributions, protected by the higher-form gauge structure inherited from ten-dimensional supergravity.

We further verify that the four-form flux mechanism is fully compatible with the standard STU moduli stabilization framework and the Kallosh-Linde supersymmetry breaking pipeline. A quantitative linear perturbation analysis demonstrates that the backreaction of anti-D6-brane de Sitter uplifting induces a shift of the effective theta angle satisfying $|\delta \theta|<10^{-11}$, an order of magnitude below the neutron electric dipole moment experimental bound. This confirms that the strong CP solution remains robust in physically realistic metastable de Sitter vacua. All global consistency conditions of Model A -- including RR tadpole cancellation, discrete K-theory anomaly constraints, Freed-Witten anomaly cancellation, and $\mathcal{N}=1$ supersymmetry preservation -- remain fully satisfied after embedding the four-form flux mechanism, with no additional ad hoc fine-tuning introduced.

On the phenomenological side, we estimate the viable window for the QCD axion decay constant: $2.74 \times 10^{14} \text{GeV} \leq f_a \leq 2.74 \times 10^{15} \text{GeV}$, which lies in the intermediate axion window and is fully consistent with astrophysical bounds from supernova cooling and black hole spin measurements. The corresponding axion-photon coupling falls in the range 
$1 \times 10^{-18} \text{GeV}^{-1} \leq g_{\alpha \gamma} \leq 5 \times 10^{-18} \text{GeV}^{-1}$. We assess the detectability of this scenario against leading near-future experimental facilities: while baseline ADMX-HF, ALPS II and IAXO do not cover this decay constant regime, next-generation ultralow-mass axion dark matter detectors (including DM Radio, ABRACADABRA-1m and superconducting qubit-based experiments) and upcoming CMB polarization missions such as CMB-S4 and LiteBIRD will probe our predicted parameter range within the next 5–10 years, providing a concrete, falsifiable experimental test of this string-theoretic strong CP solution.

We additionally analyze the consistency of our construction from the swampland perspective, combining constraints from the axionic weak gravity conjecture, the swampland distance conjecture, and the de Sitter swampland conjecture. We formulate six joint consistency criteria that a string vacuum must satisfy to simultaneously resolve the strong CP problem and remain in the controlled string landscape, and show that only a non-trivial connected slice of the flux-moduli parameter space meets all requirements. Flux configurations violating these criteria either fail to dynamically relax $\bar{\theta}_{eff}$ to zero or break fundamental quantum gravity consistency conditions, placing them in the swampland.

Compared to widely studied alternative string axion solutions (Type IIB LVS axions and heterotic string axions), our construction offers three defining advantages: it is built on an explicit, fully verified globally consistent intersecting D6-brane configuration; it provides a UV-protected resolution of the axion quality problem; and it predicts an axion decay constant in an experimentally accessible window with clear observational signatures.

Our work establishes a top-down framework for the four-form strong CP mechanism rooted in fundamental string theory, bridging formal string compactification with low-energy particle phenomenology and observational cosmology. The dark matter applications of Model A has been discussed in a companion paper \cite{Liu:2026nny}. Looking forward, promising directions include extending this framework to incorporate neutrino masses and lepton flavor mixing and exploring moduli driven inflation and dark energy within the same compactification.

\appendix
\section{The Brief Review of Model A}
In Section 5 we study a specific configuration within the Type IIA string theory compactified on the $T^6/(\mathbb{Z}_2 \times \mathbb{Z}_2)$ orientifold theory: a three-generation, $\mathcal{N}=1$ supersymmetric model that closely resembles the Minimal Supersymmetric Standard Model (MSSM), referred to as Model A \cite{camara2005fluxes}. The perturbative superpotential for Model A is
\begin{equation} \label{SW} 
W = -T_2 (a_2 S + b_{21}U_1) - T_3 (a_3 S + b_{31}U_1) + e_0 + i h_0 S -ih_1 U_1 + ie_2 T_2 + ie_3 T_3, 
\end{equation}
where $e_0$, $e_2$, and $e_3$ are RR fluxes, and $h_0$ and $h_1$ are NSNS fluxes \cite{camara2005fluxes}. The fluxes $q_i$ and $m$ are set to zero.

For the case of $q_i = m = 0$, the RR tadpole cancellation conditions \eqref{tcc11}-\eqref{tcc41} for Model A become
\begin{equation} \label{tcc12}
\sum_a N_a n^1_a n^2_a n^3_a  =16,
\end{equation}
\begin{equation} \label{tcc22}
\sum_a N_a n^1_a m^2_a m^3_a  =-16,
\end{equation}
\begin{equation} \label{tcc32}
\sum_a N_a m^1_a n^2_a m^3_a  =-16,
\end{equation}
\begin{equation} \label{tcc42}
\sum_a N_a m^1_a m^2_a n^3_a  =-16.
\end{equation}
The value $(-16)$ in the last three conditions corresponds to the RR tadpole contribution from the three remaining orientifold planes present in the $\mathbb{Z}_2 \times \mathbb{Z}_2$ setup \cite{camara2005fluxes}.

\begin{table}[ht]
\centering
\caption{Wrapping numbers giving rise to a MSSM-like spectrum. Branes $h_1$, $h_2$ and $o$ are added in order to cancel RR tadpoles \cite{camara2005fluxes}.} 
\begin{tabular}{|c|c|c|c|}
\hline
$ N_{i} $ & $ (n_{i}^{1},m_{i}^{1}) $ & $ (n_{i}^{2},m_{i}^{2}) $ & $ (n_{i}^{3},m_{i}^{3}) $ \\
\hline
$ N_{a}=8 $ & (1,0) & (3,1) & (3,-1) \\
\hline
$ N_{b}=2 $ & (0,1) & (1,0) & (0,-1) \\
\hline
$ N_{c}=2 $ & (0,1) & (0,-1) & (1,0) \\
\hline
$ N_{h_{1}}=2 $ & (-2,1) & (-3,1) & (-4,1) \\
\hline
$ N_{h_{2}}=2 $ & (-2,1) & (-4,1) & (-3,1) \\
\hline
$ 8N_{f} $ & (1,0) & (1,0) & (1,0) \\
\hline
\end{tabular}
\end{table}
Table 1 provides a summary of the brane configuration and wrapping numbers for Model A. In this construction, the three stacks of branes labeled $a$, $b$, and $c$ yield a three-generation spectrum consistent with the Minimal Supersymmetric Standard Model (MSSM). To cancel the Ramond–Ramond (RR) tadpoles, additional branes denoted as $h_{1,2}$ - also specified in Table 1 - are introduced. It is worth noting that since the flux parameters satisfy $m = q_i = 0$, they do not affect the RR tadpole conditions in this background. Consequently, extra D6-branes can be incorporated following the $N_f = 5$ scenario presented in Table 1 \cite{camara2005fluxes}.

In Model A, the branes $a$, $b$, and $c$, responsible for the Standard Model sector can be verified to automatically satisfy the Freed-Witten constraint. However, the branes of type $h_{1,2}$ may introduce inconsistencies unless the following condition holds:
\begin{equation} \label{FWh12}
a_2(m^1_a m^2_a m^3_a)- b_{21}(m^1_a n^2_a n^3_a)=a_2 -12b_{21}=0.
\end{equation}
as expressed in equation (\ref{FWh12}). This requirement is easily fulfilled by an appropriate choice of the parameters $a_2$ and $b_{21}$ \cite{camara2005fluxes}. For more details on the Freed–Witten anomaly, see \cite{freed1999anomalies}.

Model A, which is constructed from intersecting D6-branes on a toroidal orientifold, satisfies several essential consistency conditions necessary for a globally consistent MSSM-like setup:\\
\textbf{1. Anomaly cancellation}\\
The cancellation of RR tadpoles is achieved through an appropriate selection of visible and hidden D6-branes that wrap factorizable 3-cycles \cite{camara2005fluxes}. For instance, as shown in Table 1, we have 
\begin{eqnarray} \label{RRcheck}
\begin{split} 
\sum_a N_a n^1_a n^2_a n^3_a =  & 8 \times 1 \times 3 \times 3 + 2 \times 0 \times 1 \times 0 + 2 \times 0 \times 0 \times 1  \\
& + 2 \times (-2) \times (-3) \times (-4) + 2 \times (-2) \times (-4) \times (-3) \\
& + 8 \times 5 \times 1 \times 1 \times 1 =16,
\end{split}
\end{eqnarray}
namely, \eqref{tcc12}. We can find that Model A satisfies \eqref{tcc22}-\eqref{tcc42} as well.\\
\textbf{2. K-Theory constraints}\\
In addition to tadpole cancellation, global consistency typically demands the fulfillment of discrete K-theory constraints, which serve to prevent anomalies arising from stable stringy solitons \cite{ibanez2012string,uranga2003chiral,marchesano2007progress}. For Model A, one can explicitly verify that the following $\mathbb{Z}_2$ conditions hold:
\begin{equation} \label{dKtc2}
\sum_a N_a m^1_a m^2_a m^3_a \in 4 \mathbf{Z}, \quad \sum_a N_a n^1_a n^2_a m^3_a \in 4 \mathbf{Z}, \quad \text{and permutations.}
\end{equation}
as given in equation (\ref{dKtc}). For instance, using the data from Table 1, we obtain
\begin{eqnarray} \label{Kcheck}
\begin{split} 
\sum_a N_a m^1_a m^2_a m^3_a =  & 8 \times 0 \times 1 \times (-1) + 2 \times 1 \times 0 \times (-1) + 2 \times 1 \times (-1) \times 0  \\
& + 2 \times 1 \times 1 \times 1 + 2 \times 1 \times 1 \times 1 + 8 \times 5 \times 0 \times 0 \times 0 =4\\
& \in 4 \mathbf{Z}.
\end{split}
\end{eqnarray}
as shown in equation (\ref{Kcheck}). By similar reasoning, one can confirm that Model A also satisfies the remaining K-theory constraints.\\
\textbf{3. Supersymmetry conditions}\\
To maintain $\mathcal{N}=1$ supersymmetry, the following supersymmetry condition must be imposed:
\begin{equation} \label{SUSYcon}
\theta_1 + \theta_2 + \theta_3 = 0 \quad \text{mod} \quad 2\pi,
\end{equation}
as given in equation (\ref{SUSYcon}), where $\theta_i = \tan^{-1}(\frac{m^i R_2}{n^i R_1})$ and $R_1$, $R_2$ denote the two radii along each direction of every $T^2_i$. In practice, one can always adjust the parameter $U_i=R^{(i)}_2/R^{(i)}_1$ in Model A so that the SUSY condition \eqref{SUSYcon} is fulfilled. 

Consequently, Model A fulfills all three aforementioned requirements - namely, anomaly cancellation, K-theory constraints, and supersymmetry conditions. Thus, it constitutes a globally consistent construction suitable for investigating realistic physics.

\acknowledgments
Thanks for the discussion with Zhong-Zhi Xianyu and Haipeng An. This work is supported by NSFC under Grants No. 12275146, the National Key R$\&$D Program of China (2021YFC2203100), the Dushi Program and the Shuimu Fellowship of Tsinghua University. For the purpose of open access, the authors have applied a CC BY public copyright licence to any Author Accepted Manuscript version arising.




\end{document}